\documentclass[aps,prl,reprint,groupedaddress,preprintnumbers]{revtex4-1}

\usepackage{amsmath,amssymb,mathtools}
\usepackage{graphicx}
\usepackage{hyperref}
\usepackage{subfigure}
\usepackage{comment}
\usepackage{tensor}
\usepackage{cases}
\usepackage{color}
\usepackage{soul}

\begin{document}

\providecommand{\abs}[1]{\lvert#1\rvert}
\providecommand{\bd}[1]{\boldsymbol{#1}}

\preprint{SISSA 64/2016/FISI}

\title{Geometric Baryogenesis from Shift Symmetry}

\author{Andrea De Simone}
 \email{andrea.desimone@sissa.it}
\author{Takeshi Kobayashi}
 \email{takeshi.kobayashi@sissa.it}
\author{Stefano Liberati}
 \email{stefano.liberati@sissa.it}
\affiliation{SISSA and INFN Sezione di Trieste, Via Bonomea 265, 34136 Trieste, Italy}


\begin{abstract}
We present a new scenario for generating the baryon asymmetry of the universe that is induced by a Nambu--Goldstone (NG) boson. The shift symmetry naturally controls the operators in the theory, while allowing the NG boson to couple to the spacetime geometry as well as to the baryons. The cosmological background thus sources a coherent motion of the NG boson, which leads to baryogenesis. Good candidates of the baryon-generating NG boson are the QCD axion and axion-like fields. In these cases the axion induces baryogenesis in the early universe, and can also serve as dark matter in the late universe.
\end{abstract}


\maketitle

{\it Introduction.}---
The excess of matter over antimatter in our universe is
crucial for our very existence,
and is well supported by various observations. 
In particular, measurements of the cosmic
microwave background (CMB) give the ratio between
the baryons and the entropy of the universe as
$n_B / s \approx 8.6 \times 10^{-11}$~\cite{Ade:2015xua}.
However the origin of this baryon asymmetry still remains unexplained.

In this letter we present a natural framework for creating the baryon
asymmetry by a Nambu--Goldstone (NG) boson 
of a spontaneously broken
symmetry which we need not specify.
The guiding principle here is the shift symmetry of the NG boson,
or an approximate one for a pseudo Nambu--Goldstone (pNG) boson.
We argue that a NG boson coupled to various forces through
shift-symmetric operators naturally comes equipped with the basic
ingredients for a successful baryogenesis.

From the point of view of shift symmetry, linear couplings of a NG
boson to total derivatives, such as to topological terms, are not forbidden.
Thus with gauge fields, a NG boson can acquire dimension-five
operators of the form~$\phi F \tilde{F}$.
In particular with SU(2) gauge fields, such a term gives rise, through the anomaly equation, to a
coupling to the divergence of the baryon current, i.e.~$\phi \, \nabla_\mu j_B^\mu $.

On the other hand, gravity also provides a shift-symmetric
mass-dimension-five operator $\phi \, \mathcal{G}$,
with $\mathcal{G} = R^2 - 4 R_{\mu \nu } R^{\mu \nu} +
R_{\mu \nu \rho \sigma} R^{\mu \nu \rho \sigma}$
being the topological Gauss--Bonnet term. 
In an expanding universe, the Gauss--Bonnet coupling yields an
effectively linear potential for the massless NG boson
and sources a coherent time-derivative of the NG condensate.
This, through its coupling to the baryon current, shifts the
spectrum of baryons relative to that of antibaryons,
and therefore allows baryogenesis even in thermal equilibrium
when baryon number nonconserving processes occur rapidly. 
In other words, the NG boson mediates the effect of the spontaneous
breaking of Lorentz invariance
in an expanding universe
to a shift in the baryon/antibaryon spectra.

We will also show that this scenario can be realized with the QCD axion,
in which case the axion provides
the baryon asymmetry and dark 
matter in our universe, as well as solve the strong $CP$ problem.

Although the mechanism of generating the baryons by the spontaneous
breaking of Lorentz invariance (or $CPT$
symmetry~\cite{Greenberg:2002uu}) has been 
investigated in the past, our scenario is quite distinct from the
previous studies.
``Spontaneous baryogenesis''~\cite{Cohen:1987vi} is driven by a 
massive scalar derivatively coupled to the baryon current,
with a mass typically as large as $m \gtrsim 10^5\,
\mathrm{GeV}$~\cite{Kusenko:2014uta,DeSimone:2016ofp}. 
However such a scalar condensate can ruin the
subsequent cosmological expansion history.
Moreover, the spatial fluctuation of the scalar seeded during
inflation produces baryon isocurvature
perturbations~\cite{Turner:1988sq}, which are  
tightly constrained from CMB measurements.
These observations constrain the model parameters to lie within a
rather narrow window~\cite{DeSimone:2016ofp}. 
On the other hand, in our scenario the (p)NG boson is (nearly) massless.
The small mass makes the boson long-lived, and even allows the
baryon-generating pNG boson to play the role of dark matter.
The shift symmetry further suppresses the baryon 
isocurvature much below the observational bounds.

We should also remark that
the gravitational background playing an important role in our scenario
is reminiscent of ``gravitational baryogenesis''~\cite{Davoudiasl:2004gf},
which invokes a derivative coupling between the Ricci scalar and the
current,~$(\partial_\mu R ) j_B^{\mu}$.
Such a term seems somewhat {\it ad hoc} in the sense that gravity is
assumed to distinguish between matter and antimatter, however
it might arise with the aid of mediators.
Phenomenologically,
gravitational baryogenesis 
typically requires a quite high cosmic temperature,
and also a trace anomaly for the energy-momentum
tensor in order to have a non-vanishing $\partial_t R$ in a
radiation-dominated universe. 
In contrast, the cosmic temperature in our scenario
can be lowered due to the direct coupling between the NG boson and the
baryon current.
Furthermore, since $\mathcal{G}$ does not vanish during
radiation domination, our scenario need not rely on trace
anomalies. 

Let us also note the crucial difference with 
the model of~\cite{Odintsov:2016hgc} which considered a
coupling~$(\partial_\mu \mathcal{G} ) j_B^{\mu}$.
Such a term introduces higher derivative terms in the equations of
motion which can lead to ghost instabilities.
On the other hand, the $\phi \, \mathcal{G}$ coupling of the NG
boson does not yield higher derivatives,
and thus does not introduce extra degrees of freedom except for
$\phi$~itself.

{\it Baryogenesis with a NG Boson.}---
Following the above arguments,
we consider a theory of a shift-symmetric NG scalar~$\phi$
linearly coupled to the divergence
of the baryon current, as well as to the Gauss--Bonnet term, 
described by the Lagrangian
\begin{equation}
 \frac{\mathcal{L} }{\sqrt{-g}} =  
   \frac{M_p^2 }{2} R
 - \frac{1}{2}  \partial_\mu \phi \, \partial^\mu \phi
   + \frac{\phi }{f} \nabla_\mu j_B^{\mu}
   + \frac{\phi }{M} \mathcal{G}
   + \cdots.
   \label{L1}
\end{equation}
Here $f$ and $M$ are mass scales suppressing the dimension-five
operators, and $\nabla_\mu$ is a covariant derivative.
We have specified the relative sign of the two coupling terms
for simplicity; this sign at the end determines whether baryons or 
antibaryons are created.

The derivative coupling to the baryon current can originate from the
anomalous couplings to the SU(2) gauge fields
(in such a case the coupling term is effective when sphalerons are
in equilibrium~\cite{Daido:2015gqa});
alternatively, the term could directly be generated upon 
spontaneous symmetry breaking, as in the example
of~\cite{Cohen:1988kt}.
The gravitational coupling may also arise from the symmetry
breaking, as in this case $M$ would be naturally
associated to the coherence length of the NG condensate.

The NG boson may further couple to the lepton current, then the produced
lepton asymmetry can later be converted to the baryons;
for the purpose of our discussion it suffices to just display the baryon current.   
Regarding gravity, 
a mass-dimension-five Chern--Simons coupling $\phi R \tilde{R}$ also preserves
the shift symmetry of~$\phi$~\cite{Jackiw:2003pm},
however we omit this term since $ R \tilde{R}$ vanishes in a FRW universe. 
Purely from the point of view of shift symmetry, 
there can also be~$\phi \nabla^2 R$, 
or terms equivalent to this up to total derivatives.
However such terms introduce 
ghostly extra degrees of freedom, 
and thus we do not expect them to result from a symmetry breaking of a
stable theory~\cite{foot1}.

Shift-symmetric operators other than those shown are contained
in the dots in (\ref{L1}).
We consider them to 
have smaller effects on the 
$\phi$~dynamics compared to~$\phi \, \mathcal{G} / M$,
either because the coupled non-gravitational fields are not expected to
have large vacuum expectation values,
or the operators have mass-dimensions higher than five.
A pNG~$\phi$ can also obtain a (possibly temperature-dependent)
potential from some nonperturbative effects.
For the moment we assume such a potential to be negligible during
baryogenesis, until later when we discuss the possibility of $\phi$ being
an axion.
The Lagrangian of matter fields other than~$\phi$ is also
included in the dots.

Varying the Lagrangian~(\ref{L1}) in terms of $g_{\mu \nu}$
and dropping total derivatives 
gives the Einstein's equation
(if $j_B^\mu$ is a fermion current one should instead use vierbeins, however
this actually does not affect the results~\cite{DeSimone:2016ofp}),
\begin{widetext}
\begin{equation}\label{Eineq}
\begin{split}
 M_p^2 G_{\mu \nu} = T^{\phi}_{(\mu \nu)} + T^{\mathcal{G}}_{(\mu \nu)}
 + T^{\mathrm{dots}}_{(\mu \nu)},
 \qquad 
 T^\phi_{\mu \nu} = g_{\mu \nu}
  \left( -\frac{1}{2}  \partial_\rho \phi \, \partial^\rho \phi
 - \frac{\partial_\rho \phi }{f} j_B^{\rho}  \right)
  + \partial_\mu \phi \, \partial_\nu \phi
  + 2\frac{\partial_\mu \phi }{f} j_{B\nu},
\\
 T^{\mathcal{G}}_{\mu \nu} = \frac{4}{M}
  \left(
 R \nabla_\mu \nabla_\nu \phi - g_{\mu \nu} R \nabla_\rho \nabla^\rho
 \phi + 2 R_{\mu \nu} \nabla_\rho \nabla^\rho \phi
 - 4 R\indices{_\mu^\rho} \nabla_{\rho} \nabla_{\nu} \phi
 + 2 g_{\mu \nu} R^{\rho \sigma} \nabla_\rho \nabla_\sigma \phi
 - 2 R\indices{_\mu^\rho_\nu^\sigma} \nabla_\rho \nabla_\sigma \phi 
  \right).
\end{split}  
\end{equation}
\end{widetext}
Here $T_{(\mu \nu)} = \frac{1}{2} (T_{\mu \nu} + T_{\nu \mu })$,
and $T^{\mathrm{dots}}_{(\mu \nu)}$ represents 
the contributions from the dots in~(\ref{L1}). 
We also used that 
$\frac{1}{2} \mathcal{G} g_{\mu \nu } - 2 R R_{\mu \nu} +
4 R\indices{_\mu^\rho} R\indices{_\nu_\rho}
+ 4 R^{\rho \sigma}R_{\rho \mu \sigma \nu}
- 2 R_{\rho \sigma \tau \mu} R\indices{^{\rho \sigma \tau}_\nu}$
vanishes in four spacetime dimensions as a consequence of the
generalized Gauss--Bonnet theorem. 

Considering a flat FRW universe,
$ds^2 = -dt^2 + a(t)^2 d\bd{x}^2$,
the Gauss--Bonnet term is expressed in terms of the Hubble rate $H =
\dot{a} / a$ (an overdot denotes a derivative 
in terms of the cosmological time~$t$) as
\begin{equation}
 \mathcal{G} = 24 (H^4 + H^2 \dot{H}).
\end{equation}
Focusing on the homogeneous mode of the NG scalar, $\phi = \phi(t)$,
and ignoring the spatial components of the baryon current,
the Friedmann equation (i.e. $(0,0)$~component of the Einstein's
equation~(\ref{Eineq})) reads
\begin{equation}
 3 M_p^2 H^2 = \frac{\dot{\phi}^2}{2} 
  - \frac{\dot{\phi}j_B^0 }{f}  -  \frac{24 \dot{\phi} H^3 }{M}
  + T^{\mathrm{dots}}_{00}.
\label{eqFried}
\end{equation}
We suppose the right hand side to be dominated
by $T^{\mathrm{dots}}_{00}$
and that $\phi$ has a negligible effect on the cosmological expansion;
we will evaluate this condition later on.

The equation of motion of~$\phi$
that follows from the terms shown in~(\ref{L1}) is 
\begin{align}
 0 &= \nabla_\mu \nabla^\mu \phi + \frac{\nabla_\mu j_B^\mu}{f} 
 + \frac{\mathcal{G}}{M}
 \label{EoM-phi}
 \\
 &=
 - \frac{1}{a^3}
 \frac{d}{dt}
  \left\{
   a^3
   \left(
 \dot{\phi}
 - \frac{j_B^0}{f}
 - 8 \frac{H^3}{M}
   \right)
 \right\}.
 \label{EoM-homo-phi}
\end{align}
Neglecting for the moment the term with the baryon current,
the velocity of the scalar is obtained as
\begin{equation}
 \dot{\phi} = 8 \frac{H^3}{M} + \mathrm{const.} \times a^{-3}.
  \label{dotphi}
\end{equation}
On the right hand side, during inflation when $H$ is nearly constant,
the second term is expected to become negligibly tiny compared to the first one.
After inflation, $H$ redshifts as
$a^{-2}$ during radiation domination, 
and as $a^{-3/2}$ during matter domination,
hence the second term grows relative to the first.
Which term dominates during baryogenesis is set by the initial condition
of~$\dot{\phi}$, which in turn is determined by the details of spontaneous
symmetry breaking. 
Here for simplicity, we assume that the two terms are comparable 
in magnitude at the beginning of inflation;
then one can easily check that, even if the duration of inflation is
just enough to solve the horizon problem,
the first term dominates over the second
throughout the post-inflationary era until today.
Hence hereafter we ignore the $a^{-3}$~term in~(\ref{dotphi}).

Since the time component of the baryon current denotes the 
baryon number density,
i.e. $j_B^0 = n_B$,
one sees from the energy-momentum tensor~(\ref{Eineq}) that a
nonzero~$\dot{\phi}$ gives a contribution to the energy density as
$\Delta T^\phi_{00} = - n_B \dot{\phi} / f$,
hence shifts the energy level of baryons relative to that
of antibaryons.
When the particles are in thermal equilibrium,
this can be interpreted as 
a particle of type~$i$ with baryon number~$B_i$ 
obtaining an effective chemical potential of
\begin{equation}
 \mu_i = B_i \frac{\dot{\phi}}{f}  = 
  8 B_i \frac{H^3 }{f M },
\label{chem_pot}
\end{equation}
and likewise for its antiparticle but with an
opposite sign.
Thus if some baryon number violating process is in equilibrium during a
radiation-dominated epoch, a baryon asymmetry is produced.
Supposing the particles to be relativistic fermions
and ignoring their masses, the baryon density
is obtained from the Fermi--Dirac distribution as
\begin{equation}
 n_B = \sum_{i}\frac{B_i g_i \mu_i}{6} T^2
  \left\{ 1 + \mathcal{O} \left( \frac{\mu_i}{T} \right)^2 \right\},
\end{equation}
where the sum runs over all particle/antiparticle pairs~$i$ coupled to~$\phi$,
and $g_i$ counts the internal degrees of freedom of the
(anti)particle~$i$~\cite{Arbuzova:2016qfh}. 
Using the expressions for the Hubble rate 
$3 M_p^2 H^2 = (\pi^2/30) g_* T^4$
and entropy density 
$s = (2 \pi^2 / 45) g_{s*} T^3 $
during radiation domination,
the baryon-to-entropy ratio is obtained as
\begin{equation}
 \frac{n_B}{s} =
  \frac{\pi \sum_{i} B_i^2 g_i }{9 \sqrt{10} }\frac{g_*^{3/2}}{g_{s*}}
  \frac{T^5}{f M M_p^3}.
  \label{nB-s}
\end{equation}
This ratio freezes out when the baryon violating interactions fall out
of equilibrium.
Using a subscript~``dec'' to denote evaluation at the decoupling of
the baryon violating interactions
(and in particular $T_{\mathrm{dec}}$ for the decoupling temperature),
the ratio~$(n_B/s)_{\mathrm{dec}}$ should coincide
with the current value of $8.6 \times 10^{-11}$.

We have only considered the homogeneous mode of~$\phi$ in the above
discussions, however the $\phi$~field can also possess spatial fluctuations
seeded during inflation. 
Here, note that the baryon asymmetry~(\ref{nB-s}) is independent of the
field value of~$\phi$ as a consequence of the shift symmetry;
therefore the $\phi$~fluctuations do not directly propagate into baryon
isocurvature perturbations
(see also \cite{DeSimone:2016juo} where a related idea was investigated).
Still the baryon isocurvature is not strictly zero since the
$\phi$~fluctuations are not completely frozen outside the horizon
and thus yields fluctuations in~$\dot{\phi}$.
However this effect is suppressed by powers of $(k/aH)$ for a
comoving wave number~$k$,
which can easily be checked by solving the full equation of
motion~(\ref{EoM-phi}) starting from a Bunch--Davies initial
condition.
Hence the resulting baryon isocurvature is extremely small on CMB
scales which are far outside the horizon at decoupling,
being compatible with the non-observation of isocurvature.

{\it Backreaction and Consistency.}---
We now analyze the conditions under which the above calculations can be trusted.

In $\phi$'s equation of motion~(\ref{EoM-homo-phi}),
the term~$j_B^0/f$ which we have neglected represents the 
backreaction of the produced baryons on~$\phi$. 
Comparing the last two terms in~(\ref{EoM-homo-phi}) and
substituting for $j_B^0$ from the above calculations, one finds that the
baryon backreaction can be neglected upon decoupling if 
\begin{equation}
 \left|
 \left(8 \frac{H^3}{M}\right)^{-1}
 \frac{j_B^0}{f}
 \right|_{\mathrm{dec}}
  = \frac{\sum_{i} B_i^2 g_i}{6}
  \frac{T_{\mathrm{dec}}^2}{f^2}
  \ll 1.
\label{baryon-br}
\end{equation}
This is basically a requirement that the decoupling temperature should
be lower than the cutoff~$f$.
Violation of this condition would signal the breakdown of the effective
field theory. 

The effect of the $\phi$~condensate on the cosmological expansion 
can be neglected if its contribution to the Friedmann
equation~(\ref{eqFried}) is much smaller than the total density of the
universe.
This imposes, at the time of decoupling,
\begin{equation}
 \left|
 \frac{1}{3 M_p^2 H^2}
\left( 
\frac{\dot{\phi}^2}{2} 
 -  \frac{24 \dot{\phi} H^3 }{M}
	      \right)
 \right|_{\mathrm{dec}}
= \frac{160}{3} \frac{H_{\mathrm{dec}}^4}{M^2 M_p^2}
\ll 1.
\label{grav-br}
\end{equation}
Here we substituted the solution for~$\dot{\phi}$,
and also omitted $(\dot{\phi} / f) j_B^0$ as it is guaranteed to be smaller
than the other terms under~(\ref{baryon-br}).

One can also carry out a power counting estimate of the cutoff scale from
$\phi \, \mathcal{G} / M$ along the lines discussed
in~\cite{Burgess:2009ea}. Requiring the cutoff to be higher than the
relevant energy scales gives a condition somewhat similar
to~(\ref{grav-br}),
although a naive power counting may be
misleading for a Gauss--Bonnet term.
In the following discussions, we adopt~(\ref{grav-br}) as the bound on~$M$.
Let us also remark that even when $H > M$,
the condition (\ref{grav-br}) is not necessarily violated;
however, if higher dimensional gravitational couplings are
universally suppressed by~$M$ (e.g. $(R/M^2) (\partial \phi)^2 $),
then their contributions may become important.

The decoupling scale is also bounded from above by the inflation
scale~$H_{\mathrm{inf}}$, which 
is constrained by observational limits on primordial gravitational waves;
the {\it Planck} constraint~\cite{Ade:2015lrj} yields
\begin{equation}
 H_{\mathrm{dec}} < H_{\mathrm{inf}} \lesssim  9 \times 10^{13}\,
  \mathrm{GeV}.
  \label{Hinfbound}
\end{equation}

\begin{figure}
\includegraphics[width=0.75\linewidth]{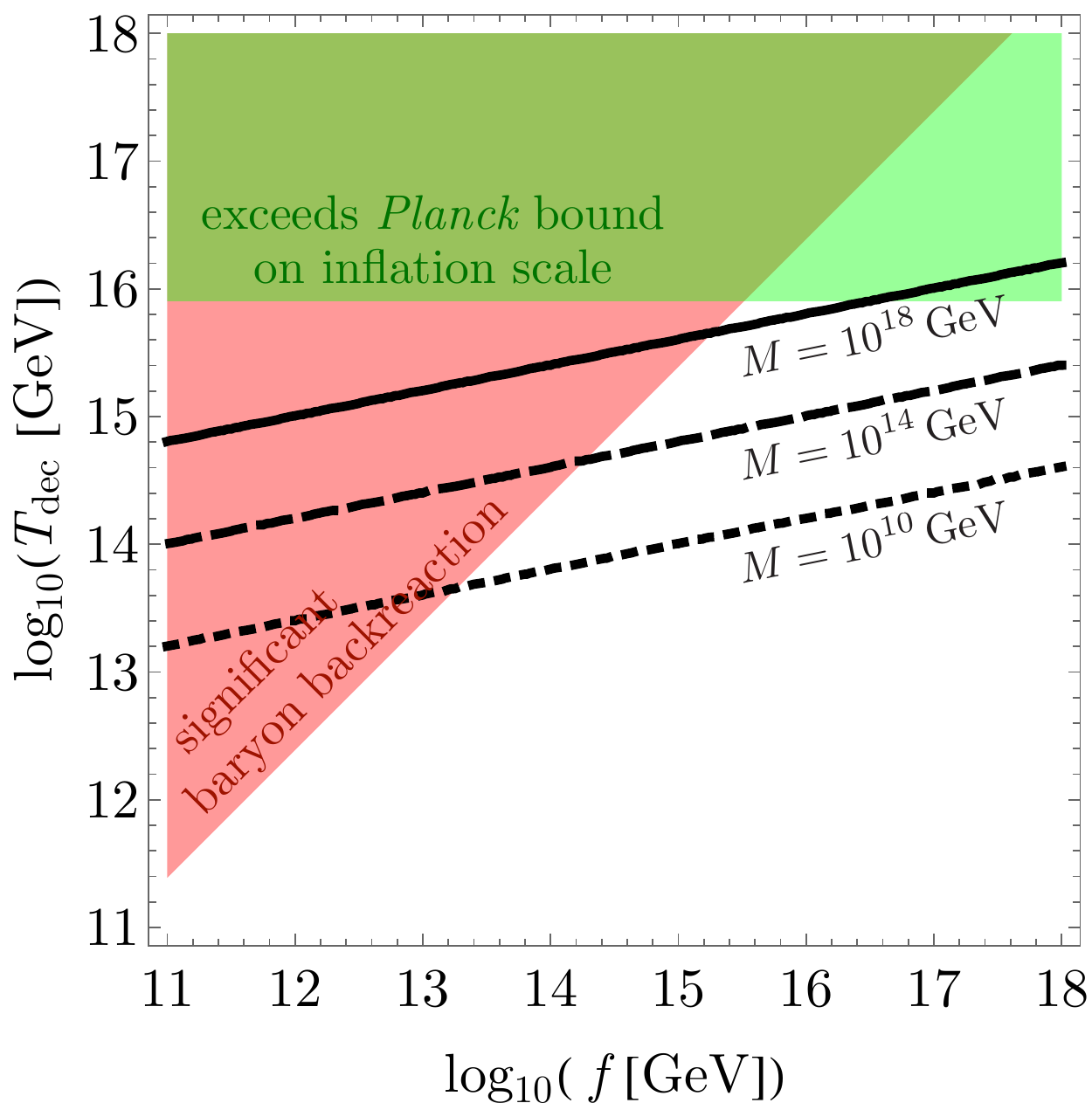}
 \caption{Parameter space in the $f$ -- $T_{\mathrm{dec}}$ plane.
 The colored regions are excluded due to significant backreaction from
 the baryons (red), and by the {\it Planck} upper bound on the inflation
 scale (green).
The allowed parameter space is shown in white. The black lines indicate where
 the right amount of baryon asymmetry is produced, for the choice of
 $M  = 10^{18}\, \mathrm{GeV}$ (solid), $10^{14}\, \mathrm{GeV}$ (dashed),
 $10^{10}\, \mathrm{GeV}$ (dotted).}
\label{fig:f-Tdec}
\end{figure}

The viable parameter space in the $f$ -- $T_{\mathrm{dec}}$ plane is
shown in Figure~\ref{fig:f-Tdec}.
Here we have chosen 
$\sum_{i} B_i^2 g_i = 1$ and 
$g_{*} (T_{\mathrm{dec}}) = g_{s*} (T_{\mathrm{dec}}) = 106.75$,
and the colored regions denote where the conditions are violated;
the red region is excluded due to significant baryon backreaction
(cf. (\ref{baryon-br})), and the green region is excluded by the {\it
Planck} bound on the inflation scale (cf. (\ref{Hinfbound})).
The black lines indicate where 
the correct amount of baryon asymmetry $(n_B/s)_{\mathrm{dec}} \approx 8.6 \times 10^{-11}$
is achieved
(cf. (\ref{nB-s})),
for $M  = 10^{18}\, \mathrm{GeV}$ (solid), $10^{14}\, \mathrm{GeV}$ (dashed),
$10^{10}\, \mathrm{GeV}$ (dotted).
For these choices of~$M$, the condition~(\ref{grav-br}) from the
gravitational backreaction is comparable to or weaker than the inflation
bound~(\ref{Hinfbound}), and thus not shown in the figure. 
For smaller~$M$, the line of $(n_B/s)_{\mathrm{dec}} \approx 8.6 \times 10^{-11}$
moves towards smaller~$T_{\mathrm{dec}}$;
the condition~(\ref{grav-br}) does not cut off the line within the
ranges of $f$ and $T_{\mathrm{dec}}$ shown in the figure,
however for $M \lesssim 10^9\, \mathrm{GeV}$, the allowed values for
$H_{\mathrm{dec}}$ exceed $M$ and thus higher dimensional gravitational
operators may become relevant. 

Further constraints can be imposed on the parameter space depending on
the nature of the NG boson. 
Let us see this directly in the following examples.

{\it QCD Axion and Axion-Like Fields.}---
Here we discuss the possibility that $\phi$ is the QCD
axion~\cite{Peccei:1977hh} which provides a solution to the strong $CP$ problem.
Then in addition to the linear potential sourced by the Gauss--Bonnet
coupling,
the axion obtains a periodic potential from non-perturbative QCD effects as
\begin{equation}
 V_{\mathrm{QCD}}(\phi, T) = m(T)^2 f_a^2
  \left\{
 1 - \cos \left(\frac{\phi }{f_a}\right)
  \right\}.
\label{axion-pot}
\end{equation}
Here $f_a$ is the axion decay constant,
and the temperature-dependent mass is 
\begin{equation}
  m(T) \approx 
 \begin{dcases}
     0.1 \times m_a \left(\frac{\Lambda_{\mathrm{QCD}}}{T}\right)^4
       & \text{for $T \gg \Lambda_{\mathrm{QCD}}$,} \\
     m_a
       & \text{for $T \ll \Lambda_{\mathrm{QCD}}$,}
 \end{dcases}
 \label{axion_mass}
\end{equation}
with
$ m_a \approx 6 \times 10^{-6}\, \mathrm{eV} 
  \left( 10^{12}\, \mathrm{GeV} / f_a  \right)$,
and $\Lambda_{\mathrm{QCD}} \approx 200 \, \mathrm{MeV}$.
Focusing on the field range $\abs{\phi} \lesssim f_a$, 
then comparison of 
$V_{\mathrm{QCD}} \simeq \frac{1}{2} m(T)^2 \phi^2 $
with the Gauss--Bonnet coupling $\phi \, \mathcal{G} / M$ in a
radiation-dominated universe shows 
that the latter dominates over the former at temperatures
\begin{equation}
 T \gtrsim 10^3\, \mathrm{GeV}
  \left( \frac{\abs{\theta(T)} M}{f_a} \right)^{1/16},
\end{equation}
where we used $\theta \equiv \phi / f_a$.
As the right hand side depends weakly on $\theta M/f_a$,
we see that as long as $T_{\mathrm{dec}} \gtrsim 10^3\, \mathrm{GeV}$
the QCD effect is negligible during baryogenesis.
On the other hand,
the Gauss--Bonnet coupling has become negligible by
the time the axion starts oscillating along its QCD potential,
which typically occurs at $T_{\mathrm{osc}} \sim 1 \, \mathrm{GeV}$.
In particular, the shift of the axion potential minimum today
due to the Gauss--Bonnet coupling is as small as
\begin{equation}
 \Delta \theta_{0} = \frac{\mathcal{G}_0}{f_a m_a^2 M} \sim
  10^{-162} \frac{f_a}{M},
\end{equation}
which
(unless for an extremely tiny~$M$)
is much smaller than the observational bound 
$\abs{\theta_0} \lesssim 10^{-10}$
from limits on the neutron electric dipole moment~\cite{Baker:2006ts}.
Thus the baryon-generating axion solves the strong $CP$
problem.

However the QCD axion~$\phi$ may overclose the universe, as
its abundance relative to cold dark matter (CDM) is given
as~\cite{Turner:1985si} 
\begin{equation}
 \frac{\Omega_\phi}{\Omega_{\mathrm{CDM}}}
  \sim \theta_{\mathrm{osc}}^2 \left( \frac{f_a}{10^{12} \,
				\mathrm{GeV}} \right)^{7/6},
\label{OemgaQCDaxion}
\end{equation}
where $\theta_{\mathrm{osc}}$ is the field value at the onset of
the axion oscillations. 
If $f_a = f$, and taking for instance the allowed values on the black
lines in Figure~\ref{fig:f-Tdec},
then the axion is long-lived and $\Omega_\phi$ can exceed unity.
One way to avoid this is by fine-tuning the misalignment~$\theta_{\mathrm{osc}}$
to a tiny value (perhaps from anthropic reasoning).
However the necessary fine-tuning is actually more severe when taking into
account the axion isocurvature perturbations~\cite{Linde:1985yf};
in order for the total CDM isocurvature to be below the CMB limit~\cite{Ade:2015lrj},
the axion can constitute only a small fraction of the entire CDM.
Moreover, since the axion field evolves in the early times due to the
Gauss--Bonnet coupling, this field excursion should also be taken into
account upon tuning the initial field value.

Alternatively, $M$ could take a low value,
provided that higher dimensional gravitational couplings are somehow
suppressed.
Then, for instance, $ M \lesssim 10^5\, \mathrm{GeV}$ allows baryogenesis
without significant backreaction
with $f_a \sim f \sim 10^{12}\, \mathrm{GeV}$;
and without fine-tuning the alignment, 
i.e. $\theta_{\mathrm{osc}} \sim 1$,
the QCD axion can generate the baryon asymmetry as well as constitute the
entire CDM. 
For these parameters, the CDM isocurvature can also be 
consistent with observational limits.

We also comment on the possibility of $\phi$ being one of the
axion-like fields arising from string theory
compactifications~\cite{Svrcek:2006yi}.
In the simplest case, such a field is described by a periodic 
potential~(\ref{axion-pot}) with a constant mass~$m$;
then its abundance is computed as
\begin{equation}
 \frac{\Omega_\phi }{\Omega_{\mathrm{CDM}}} \sim
  \theta_{\mathrm{osc}}^2 \left( \frac{f_a}{10^{17} \, \mathrm{GeV}} \right)^2
  \left( \frac{m}{10^{-22}\, \mathrm{eV}} \right)^{1/2}.
\label{Oemgaaxion-like}
\end{equation}
For example, with $\theta_{\mathrm{osc}} \sim 1$,
$ f_a \sim f \sim 10^{17}\, \mathrm{GeV}$, and
$ m \sim 10^{-22}\, \mathrm{eV}$,
the axion-like~$\phi$ can serve as CDM and
generate the baryons, cf. Figure~\ref{fig:f-Tdec}.
One can also check that if further $M \lesssim 10^{14}\, \mathrm{GeV}$,
the corresponding decoupling temperature allows
inflation scales that give CDM isocurvature
below the current limit. 
Such an ultralight axion CDM is also interesting from the point of view that 
it can produce distinct signatures on small-scale
structures~\cite{Hu:2000ke}.

{\it Discussion.}---
Without some extra symmetries, 
there is no {\it a priori} reason to forbid a NG boson from acquiring
shift-symmetric couplings to other fields. 
While most coupled fields do not induce coherent effects,
the background gravitational field of an expanding universe gives rise to
a coherent motion of the NG boson.
We have shown that this leads to the creation of a net baryon asymmetry
of the universe. 
Good candidates for the baryon-generating NG boson are the
axion(-like) fields.
This raises the intriguing possibility
that an axion could induce
baryogenesis in the early universe, then serve as cold dark matter in
the later universe
(and further solve the strong $CP$ problem if it is the QCD axion).

Let us comment on the observable consequences of our scenario.
Theories of a scalar coupled to the Gauss--Bonnet term are known to
evade no-hair theorems for black holes~\cite{Benkel:2016rlz},
which may be tested by gravitational wave observations.
We also note that if $M$ is not far from~$M_p$, 
the corresponding high decoupling temperature implies a high inflation
scale, yielding primordial gravitational waves that could be observed by
upcoming experiments. 
Furthermore, couplings between the time-dependent NG boson and
parity violating terms such as $F\tilde{F}$ may leave signatures in cosmological
observations~\cite{Lue:1998mq,foot2}.
It would also be interesting to study the experimental implications of
the required baryon violating interactions.

\begin{acknowledgments}
We would like to thank Stefano Bertolini, Latham Boyle, Shunichiro
Kinoshita, Marco Letizia, Shuntaro Mizuno, and Thomas Sotiriou for
helpful conversations. 
T.K. acknowledges support from the INFN INDARK PD51 grant. S.L. wishes to acknowledge the John Templeton Foundation for the supporting grant \#51876.
\end{acknowledgments}



\begin{thebibliography}{99}

\bibitem{Ade:2015xua} 
  P.~A.~R.~Ade {\it et al.} [Planck Collaboration],
  Astron.\ Astrophys.\  {\bf 594}, A13 (2016)
  [arXiv:1502.01589 [astro-ph.CO]].

\bibitem{Greenberg:2002uu}
	In local effective field theories  the  violation of $CPT$
	symmetries implies that of local Lorentz invariance, while
	Lorentz breaking operators can be either $CPT$ odd or
	even; see 
  O.~W.~Greenberg,
  Phys.\ Rev.\ Lett.\  {\bf 89}, 231602 (2002)
  [hep-ph/0201258].

\bibitem{Cohen:1987vi} 
  A.~G.~Cohen and D.~B.~Kaplan,
  Phys.\ Lett.\ B {\bf 199}, 251 (1987).

\bibitem{Kusenko:2014uta} 
  A.~Kusenko, K.~Schmitz and T.~T.~Yanagida,
  Phys.\ Rev.\ Lett.\  {\bf 115}, no. 1, 011302 (2015)
  [arXiv:1412.2043 [hep-ph]].

\bibitem{DeSimone:2016ofp} 
  A.~De Simone and T.~Kobayashi,
  JCAP {\bf 1608}, no. 08, 052 (2016)
  [arXiv:1605.00670 [hep-ph]].

\bibitem{Turner:1988sq} 
  M.~S.~Turner, A.~G.~Cohen and D.~B.~Kaplan,
  Phys.\ Lett.\ B {\bf 216}, 20 (1989).
	
\bibitem{Davoudiasl:2004gf} 
  H.~Davoudiasl, R.~Kitano, G.~D.~Kribs, H.~Murayama and P.~J.~Steinhardt,
  Phys.\ Rev.\ Lett.\  {\bf 93}, 201301 (2004)
  [hep-ph/0403019].	

\bibitem{Odintsov:2016hgc} 
  S.~D.~Odintsov and V.~K.~Oikonomou,
  Phys.\ Lett.\ B {\bf 760}, 259 (2016)
  [arXiv:1607.00545 [gr-qc]].

\bibitem{Daido:2015gqa} 
  R.~Daido, N.~Kitajima and F.~Takahashi,
  JCAP {\bf 1507}, no. 07, 046 (2015)
  [arXiv:1504.07917 [hep-ph]];
  B.~Shi and S.~Raby,
  Phys.\ Rev.\ D {\bf 92}, no. 8, 085008 (2015)
  [arXiv:1507.08392 [hep-ph]].

\bibitem{Cohen:1988kt} 
  A.~G.~Cohen and D.~B.~Kaplan,
  Nucl.\ Phys.\ B {\bf 308}, 913 (1988).

\bibitem{Jackiw:2003pm} 
  R.~Jackiw and S.~Y.~Pi,
  Phys.\ Rev.\ D {\bf 68}, 104012 (2003)
  [gr-qc/0308071].
	
\bibitem{foot1} 
	The extra degrees of freedom could be made stable by further modifying
	the gravity theory, or when taking into account all 
	higher dimensional operators, however we do not pursue these
	possibilities in this letter.

\bibitem{Arbuzova:2016qfh}
	Alternatively, one can derive the baryon asymmetry by studying
	the Hamiltonian, as was performed in e.g.	
  E.~V.~Arbuzova, A.~D.~Dolgov and V.~A.~Novikov,
  Phys.\ Rev.\ D {\bf 94}, no. 12, 123501 (2016)
  [arXiv:1607.01247 [astro-ph.CO]].

\bibitem{DeSimone:2016juo} 
  A.~De Simone and T.~Kobayashi,
  JCAP {\bf 1702}, no. 02, 036 (2017)
  [arXiv:1610.05783 [hep-ph]].
	
\bibitem{Burgess:2009ea} 
  C.~P.~Burgess, H.~M.~Lee and M.~Trott,
  JHEP {\bf 0909}, 103 (2009)
  [arXiv:0902.4465 [hep-ph]];
%
  M.~P.~Hertzberg,
  JHEP {\bf 1011}, 023 (2010)
  [arXiv:1002.2995 [hep-ph]];
%
  F.~Bezrukov, A.~Magnin, M.~Shaposhnikov and S.~Sibiryakov,
  JHEP {\bf 1101}, 016 (2011)
  [arXiv:1008.5157 [hep-ph]].
	
\bibitem{Ade:2015lrj} 
  P.~A.~R.~Ade {\it et al.} [Planck Collaboration],
  Astron.\ Astrophys.\  {\bf 594}, A20 (2016)
  [arXiv:1502.02114 [astro-ph.CO]].
	
\bibitem{Peccei:1977hh} 
  R.~D.~Peccei and H.~R.~Quinn,
  Phys.\ Rev.\ Lett.\  {\bf 38}, 1440 (1977);
  S.~Weinberg,
  Phys.\ Rev.\ Lett.\  {\bf 40}, 223 (1978);
  F.~Wilczek,
  Phys.\ Rev.\ Lett.\  {\bf 40}, 279 (1978).

\bibitem{Baker:2006ts} 
  C.~A.~Baker {\it et al.},
  Phys.\ Rev.\ Lett.\  {\bf 97}, 131801 (2006)
  [hep-ex/0602020].
	
\bibitem{Turner:1985si} 
  M.~S.~Turner,
  Phys.\ Rev.\ D {\bf 33}, 889 (1986).

\bibitem{Linde:1985yf} 
  A.~D.~Linde,
  Phys.\ Lett.\  {\bf 158B}, 375 (1985);
%
  D.~Seckel and M.~S.~Turner,
  Phys.\ Rev.\ D {\bf 32}, 3178 (1985);
%
  D.~H.~Lyth,
  Phys.\ Lett.\ B {\bf 236}, 408 (1990);
%
  T.~Kobayashi, R.~Kurematsu and F.~Takahashi,
  JCAP {\bf 1309}, 032 (2013)
  [arXiv:1304.0922 [hep-ph]].
	
\bibitem{Svrcek:2006yi} 
  P.~Svrcek and E.~Witten,
  JHEP {\bf 0606}, 051 (2006)
  [hep-th/0605206];
%
  A.~Arvanitaki, S.~Dimopoulos, S.~Dubovsky, N.~Kaloper and J.~March-Russell,
  Phys.\ Rev.\ D {\bf 81}, 123530 (2010)
  [arXiv:0905.4720 [hep-th]].

\bibitem{Hu:2000ke} 
  W.~Hu, R.~Barkana and A.~Gruzinov,
  Phys.\ Rev.\ Lett.\  {\bf 85}, 1158 (2000)
  [astro-ph/0003365];
%
  L.~Hui, J.~P.~Ostriker, S.~Tremaine and E.~Witten,
  Phys.\ Rev.\ D {\bf 95}, no. 4, 043541 (2017)
  [arXiv:1610.08297 [astro-ph.CO]].	

\bibitem{Benkel:2016rlz} 
  R.~Benkel, T.~P.~Sotiriou and H.~Witek,
  Class.\ Quant.\ Grav.\  {\bf 34}, no. 6, 064001 (2017)
  [arXiv:1610.09168 [gr-qc]].
	
\bibitem{Lue:1998mq} 
  A.~Lue, L.~M.~Wang and M.~Kamionkowski,
  Phys.\ Rev.\ Lett.\  {\bf 83}, 1506 (1999)
  [astro-ph/9812088].

\bibitem{foot2}
	One may also expect gauge field excitation through operators
	of the form $(\phi / f) F\tilde{F}$, however this effect is
	insignificant at decoupling as $\abs{\dot{\phi}/H
	f}_{\mathrm{dec}} < 1$ is satisfied in the parameter
	space of Figure~\ref{fig:f-Tdec}.
	Couplings between $\phi$ and graviton excitations in the
	Gauss--Bonnet term are further suppressed by powers of~$M_p$.

	
\end{thebibliography}
\end{document}